\documentclass[runningheads,a4paper]{llncs}
\usepackage{longtable}

\usepackage{courier}
\usepackage{listings}
\usepackage{multirow}
\usepackage{tablefootnote}

\usepackage{pdflscape}
\usepackage{afterpage}

\makeatletter
\newcommand{\@chapapp}{\relax}%
\makeatother

\usepackage[title,toc,titletoc,header]{appendix}

\lstdefinestyle{generalstyle}{
 belowcaptionskip=1\baselineskip,
 xleftmargin=\parindent,
 basicstyle=\scriptsize\ttfamily,
 breaklines=true,
 showstringspaces=false
}

\def\Title{A Study of Concurrency Bugs and Advanced Development Support for Actor-based Programs}
\def\PdfTitle{\Title}
\def\PdfSubject{Development Support for Actor-based Programs}
\def\PdfAuthors{Carmen Torres Lopez, Stefan Marr, Hanspeter Moessenboeck, Elisa Gonzalez Boix}
\def\PdfKeywords{Concurrency, Bug, Debugging, Analysis, Testing, Actor-based languages, Event-loop concurrency, Truffle}

\usepackage[utf8]{inputenc}
\usepackage{fixltx2e} %
\usepackage{ifpdf}
\usepackage{graphicx}

\newif\ifhtml
\ifpdf
\htmlfalse
\else
\htmltrue
\fi

\usepackage{collab}  %

\ifpdf
    \usepackage[bookmarks=true,bookmarksnumbered=true,
        pdfpagemode={UseOutlines},plainpages=false,pdfpagelabels=true,
        colorlinks=true,linkcolor={black},citecolor={black},urlcolor={black},
        pdftitle={\PdfTitle},%
        pdfsubject={\PdfSubject},%
        pdfauthor={\PdfAuthors},%
        pdfkeywords={\PdfKeywords},
        pdftex]{hyperref}

\else
    \ifhtml
        \usepackage[        colorlinks=true,linkcolor={black},citecolor={black},urlcolor={black},
        tex4ht]{hyperref}
    \else
        \usepackage[        colorlinks=true,linkcolor={black},citecolor={black},urlcolor={black}]{hyperref}
    \fi
\fi

\newcommand{\ie}{i.e.\xspace}
\newcommand{\eg}{e.g.\xspace}

\usepackage{letltxmacro}
\newcommand{\code}[1]{\lstinline!#1!}

\usepackage{ifthen}

\usepackage[nameinlink]{cleveref}
\Crefformat{lstlisting}{#2Lst.\,#1#3}
\crefformat{lstlisting}{#2lst.\,#1#3}

\usepackage[square,numbers]{natbib}
\usepackage[utf8]{inputenc}
\usepackage{fixltx2e} % for \textsubscript
\usepackage{ifpdf}
\usepackage{graphicx}

\newif\ifhtml
\ifpdf
\htmlfalse
\else
\htmltrue
\fi

\usepackage{collab}  % [hidecomments]
\usepackage{letltxmacro}
\usepackage{ifthen}
\newcommand{\citeurl}[5]{%
#1\footnote{\emph{#2}%
          \ifthenelse{\equal{#3}{}}
                     {}%          then i.e. empty #3
                     {, #3}% else i.e. non-empty #3
          \ifthenelse{\equal{#4}{}}
                     {}%          then i.e. empty #4
                     {, access date: #4}% else i.e. non-empty #4
          , \url{#5}}}

\usepackage[nameinlink]{cleveref}
\Crefformat{lstlisting}{#2Lst.\,#1#3}
\crefformat{lstlisting}{#2lst.\,#1#3}

\begin{document}

\mainmatter  %

\title{\Title}
\titlerunning{Concurrency Bugs and Development Support for Actor-based Programs}

\author{
Carmen Torres Lopez\inst{1}\and
Stefan Marr\inst{2}\and
Elisa Gonzalez Boix\inst{1}\and
Hanspeter Mössenböck\inst{2}}
\institute{
Vrije Universiteit Brussel, Pleinlaan 2, 1050, Brussel, Belgium\and
Johannes Kepler University, Linz, Austria}
\authorrunning{Torres Lopez et al.}

\toctitle{toctitle TODO}
\tocauthor{tocauthor TODO}

\maketitle
\begin{abstract}
The actor model is an attractive foundation for developing concurrent applications because actors are isolated concurrent entities that communicate through asynchronous messages and do not share state. 
Thereby, they avoid concurrency bugs such as data races, but are not immune to concurrency bugs in general.  

This study taxonomizes concurrency bugs in actor-based programs reported in literature.
Furthermore, it analyzes the bugs to identify the patterns causing them as well as their observable behavior.
Based on this taxonomy, we further analyze the literature and find that current
approaches to static analysis and testing focus on communication deadlocks and message protocol violations. However, they do not provide solutions
to identify livelocks and behavioral deadlocks.

The insights obtained in this study can be used to improve debugging support for actor-based programs with new debugging techniques to identify the root cause of complex concurrency bugs.

\end{abstract}

\section{Introduction}

With the widespread use of multicore systems, even in everyday phones, concurrent programming has become mainstream.
However, concurrent programming is known to be hard and error-prone.
Unlike traditional sequential programs, concurrent programs often exhibit \emph{non-deterministic} behavior which makes it difficult to reason about their behavior.
Many bugs involving concurrent entities, \eg processes, threads, actors\cite{Agha:1985}, manifest themselves only in rare execution traces.
Identifying and analyzing concurrency bugs is thus an arduous task, perhaps even an art. 

When studying techniques to support the development of complex concurrent programs, our first research question is what types of concurrency bugs appear in such programs.
The answer to this question depends on the concurrency model in which the program is written.
Most existing studies about concurrency bugs focus on thread-based concurrency\cite{Artho:2003, Peierls:2005, Brito:2010, Lu:2008, Martin:2012, Leesatapornwongsa:2016, Abbaspour:2016, Abbaspour:2016b}.

The established frame of reference, however, does not directly apply to other concurrency models which are not based on a shared memory model such as the actor model, communicating sequential processes (CSP), etc.
In this paper we study concurrency bugs in message passing concurrent software, in particular, in actor-based programs. %

The actor model is attractive for concurrent programming because it avoids by design some concurrency bugs associated with thread-based programs.
Since actors do not share mutable state, programs cannot exhibit memory-level \emph{race conditions} such as data races.
In addition to that, deadlocks can be avoided if communication between actors is solely based on asynchronous message passing.
However, this does not mean that programs are inherently free from concurrency issues.

This paper surveys concurrency bugs in the literature on actor-based programs and aims to answer three research questions:
(1) which kind of concurrency bugs can be avoided by the actor model and its variants, 
(2) what kind of patterns cause concurrency bugs in actor programs,
and (3) what is the observable behavior in the programs that have these bugs?

To provide a common frame of reference to distinguish different types of concurrency bugs that appear in actor-based programs, we propose a
taxonomy of concurrency bugs in actor-based programs (in Section~\ref{sec:bugs}).
The taxonomy aims to establish a conceptual framework for concurrency bugs that facilitates communication amongst researchers.
It is also meant to help practitioners in developing, testing, debugging, or even statically analyzing programs to identify the root cause of concurrency bugs by offering more information about the types of bugs and their observable properties.

Based on our taxonomy of bugs, we analyze actor literature that reports concurrency bugs and map them to the proposed classification.
Furthermore, we identify which types of bugs have been addressed in literature
so far, and which types have been studied less.

The contributions of this paper are:
\begin{itemize}
\item A systematic study of concurrency bugs in actor-based programs based on a literature review. To the best of our knowledge it is the first taxonomy of bugs in the context of actor-based concurrent software. 

\item An analysis of the patterns and observable behaviors of concurrency bugs found in different actor-based programs.

\item A review of the state of the art in static analysis, testing, debugging, and visualization of actor-based programs to identify open research issues.

\end{itemize}

\section{Terminology and Background Information}
Before we delve into the classification of concurrency bugs in actor-based programs, we discuss the terminology used in this paper and the basic concepts on actor-based programs and concurrency issues.

Since the actor model was first proposed by Hewitt et al. \cite{Hewitt:1973}, several variations of it emerged.
Based on De Koster et al. \cite{DeKoster:2016:YAT}, we distinguish three variants in addition to the classic actor model:
\emph{active objects} (\eg ABCL \cite{Yonezawa:1986}, AmbientTalk/1\cite{Dedecker:2006}),
\emph{processes} (\eg Erlang \cite{Armstrong:1993}, Scala)
and \emph{communicating event-loops} (\eg E\cite{Miller:2005}, AmbientTalk/2\cite{Vancutsem:2007}, JavaScript). 
In all these variants, concurrency is introduced by actors.
All actors communicate with one another by means of \emph{messages}.
Messages are stored in a \emph{mailbox}.
Each actor has a thread of execution, which perpetually processes one message at a time from the mailbox.
The processing of one message by an actor defines a \emph{turn}.
Each actor has a \emph{behavior} associated that defines how the actor processes messages.
The set of messages that an actor knows how to process in a certain turn denotes the \emph{interface} of the actor's behavior.
Actors can store state which can only be accessed or mutated by the actor itself.
In other words, actors have exclusive access to their mutable state. 

A \emph{concurrency bug} is a failure related to the interactions among different concurrent entities of a system.
Following Avizienis's terminology\cite{Avizienis:2004}, a \emph{failure} is an event that occurs when the services provided by a system deviate from the ones it was designed for. %
The discrepancy between the observed behavior and the theoretically correct behavior of a system is called an \emph{error}. %
Hence, an error is an event that may lead to a failure.
Finally, a \emph{fault} is an incorrect step in a program which causes an error (\eg the cause of a message transmission error in a distributed system may be a broken network cable).
A fault is said to be \emph{active} when it causes an error, and \emph{dormant} when is present in a system but has not yet manifested itself as an error.
Throughout this paper, we use the terms concurrency bug and issue interchangeably.

Although actors were originally designed to be used in open distributed environments, they can be used on a single machine, e.g. in multicore programming.
This paper analyses concurrency bugs that appear in actor-based programs used in either concurrent or distributed systems.
However, bugs that are only observable in distributed systems (e.g. due to network failures) are out of the scope of this paper.

\section{Classification of Concurrency Bugs in Actor-based Programs}
\label{sec:bugs}
While there is a large number of studies for concurrency bugs in thread-based programs, there are only few studies on bugs in the context of message passing programs.
Zhang et al. \cite{Zhang:2015} study bug patterns, manifestation conditions, and bug fixes in three open source applications that use message passing.
In this context, literature typically uses general terms to refer a certain issue, for example ordering problems\cite{Long:2016}.
For actor-based programs however, there is so far no established terminology for concurrency bugs. 

This section introduces a taxonomy of concurrency bugs for the actor model derived from bugs reported in literature and from our own experience with actor languages.
\Cref{tab:taxonomy} first summarizes the well-known terminology for thread-based programs from literature, and then introduces our proposed terminology for concurrent bugs in actor-based programs.
Our overall categorization starts out from the distinction of shared-memory concurrency bugs in literature, which classifies bugs in two general categories:
lack of progress issues and race conditions.

\afterpage{%
\begin{landscape}
\begin{table}
\centering
\begin{tabular}{|p{2cm}|p{2.5cm}|p{4cm}|p{10cm}|}
  \hline
    \multicolumn{1}{|c|}{Concurrency Model} &\multicolumn{2}{|c|}{Category of Concurrency Bugs}& \multicolumn{1}{|c|}{Bug Definition}  \\\hline
      
\multirow{6}{*}{Threads}& 
	\multirow{2}{*}{Lack of Progress} & Deadlock & condition in a system where two or more threads are blocked forever waiting for another thread to do something\cite{Prasad:2015}.\\
	\cline{3-4}
	 & & Livelock & condition in which two or more threads while not blocked cannot make further progress\cite{Peierls:2005}.
	 \\
	 \cline{2-4}
	& \multirow{3}{*} {Race Condition} & Data race & special case of race condition that occurs when two threads access the same data and at least one of them writes the data \cite{Abbaspour:2016}.\\
	\cline{3-4}
	 & & Bad interleaving (also know as high-level data race\cite{Artho:2003}, atomicity violation\cite{Abbaspour:2016}) & occurs when the program exposes an inconsistent intermediate state due to the overlapping execution of two threads\cite{Prasad:2015}.\\
	 \cline{3-4}
     &  & Order violation & occurs when the expected order of execution of at least two memory accesses is not respected\cite{Abbaspour:2016}.\\
    \hline
\multirow{5}{*}{Actors}& 
	\multirow{3}{*}{Lack of Progress} & Communication deadlock & condition in a system where two or more actors are blocked forever waiting for each other to do something. \\
	\cline{3-4}
	 & & Behavioral deadlock & condition in a system when two or more actors are not blocked but wait on each other for a message to be able to progress, \ie the message to complete the next step is never sent.\\
	 \cline{3-4}
	 & & Livelock & condition similar to a deadlock in which two or more actors are not able to make progress but they continuously change their state.  \\
	     
	  \cline{2-4}
	& \multirow{3}{4em} {Message Protocol Violation} & {Message order violation} & condition in which the order of exchanging messages of two or more actors is not consistent with the intended \emph{protocol} of an actor.\\ 
     \cline{3-4}
	 \cline{3-4}
	& & {Bad message interleaving} & occurs when a message is processed between two messages which are intended to be processed one after the other.\\
	 \cline{3-4}
    & & {Memory inconsistency} & occurs when different actors have inconsistent views of shared resources. The effects of the turn that modifies a conceptually shared resource, may not be visible to other actors which also alter the same resource. \\
    \hline    
    
\end{tabular}
\caption{Taxonomy of concurrency bugs}
\label{tab:taxonomy}
\end{table}
\end{landscape}
}%

Depending on the guarantees provided by a specific actor model, programs may
be subject to different concurrency bugs. Therefore, not all concurrency
bugs are applicable to all actor variants. 
In the rest of the section we define each type of bug, and detail in which variants it cannot be present.

\subsection{Lack of Progress Issues}
Two different kinds of conditions can lead to a lack of progress in an actor-based program: deadlocks and livelocks.
However, these issues manifest themselves differently in actor-based programs compared to thread-based programs.

\subsubsection{Communication Deadlock.}
A communication deadlock is a condition in a system where two or more actors are blocked forever waiting for each other to do something. 
This condition is similar to traditional deadlocks known from thread-based programs.
We base the terminology on the work of \cite{Christakis:2011} in Erlang concurrency bugs.

Communication deadlocks can  \emph{only} occur in variants of the actor model that feature a blocking \code{receive} operation.
This is common in variants of the actor model based on processes.
Examples of such actor systems include Erlang and the Scala Actors framework\cite{Haller:2009}.
A communication deadlock manifests itself when an actor only has messages in its inbox that cannot be received with the currently active \code{receive} statement.
\Cref{lst:pingpong} shows a communication deadlock example in Erlang\cite{Christakis:2011}.
The \emph{fault} is in \cref{ln:rcvr}, where the \code{pong} process is blocked because it is waiting for a message that is never sent by the \code{ping} process.
Instead the \code{ping} process returns \code{ok}.

\begin{lstlisting}[caption=Communication deadlock example in Erlang (from \cite{Christakis:2011}). \Cref{ln:rcvr} has a blocking \code{receive} causing the \code{pong} process to deadlock because the expected message is never sent. ,label=lst:pingpong, language=erlang,float=thpb]
play() ->
  Ping = spawn(fun ping/0),
  spawn(fun() -> pong(Ping) end).

ping() ->
  receive
    pong_msg -> ok
  end.

pong(Ping) ->
  Ping ! pong_msg,
  receive @*\label{ln:rcvr}*@
    ping_msg -> ok
  end.
\end{lstlisting}

\subsubsection{Behavioral Deadlock.}

A behavioral deadlock happens when two or more actors \emph{conceptually} wait for each other because the message to complete the next step in an algorithm is never sent. In this case, no actor is necessarily suspended or otherwise unable to receive messages. 
We call this situation a behavioral deadlock, because the mutual waiting prevents local progress. 
However, these actors might still process messages from other actors.
Since actors do not actually block, detecting behavioral deadlocks can be harder than detecting deadlocks in thread-based programs.

We illustrate a behavioral deadlock in an implementation of the dining philosophers concurrency problem written in Newspeak\cite{Bracha:10:NS} which is shown in \Cref{lst:philosophers}.
The behavioral deadlock has the effect that some philosophers cannot eat (as they never acquire two consecutive forks), preventing global progress.
\Cref{ln:rfid} shows that the left fork has the same value as the \code{id} of the philosopher, but for the right fork the program computes its value. %
For example, philosopher 1 will eat with fork 1 and 2 and so on. 
The \emph{error} occurs when the philosopher puts down its forks: the right fork gets a wrong value (\cref{ln:nforks}) because the implementation swapped \code{numForks} and \code{leftForkId} variables. This programming mistake is the \emph{fault} that causes fork 2 and 4 to be always taken. 
Consequently, there is no global progress since philosopher 2 and 4 never eat and philosopher 1 and 3 eat only once.
Philosopher 5 can always eat showing local progress, however. 

\begin{lstlisting}[caption={Behavioral deadlock example of a dining philosopher implementation. \Cref{ln:nforks} calculates \texttt{rightForkId} incorrectly, preventing the philosophers from eating.},label=lst:philosophers,float=thpb] 
class PhilosopherActor new: id rounds: rounds 
    counter: aCounter arbitrator: arbitrator = (
  (* ... *)
  public start = (
    arbitrator <-: pickUpForks: self id: id. @*\label{ln:start}*@
  )
)  
class ArbitratorActor new: numForks resolver: resolver = (
  (* ... *)
  public pickUpForks: philosopher id: leftForkId = (
    | rightForkId |
    rightForkId := 1 + (leftForkId % numForks). @*\label{ln:rfid}*@
    ((forks at: leftForkId) or: [forks at: rightForkId])
      ifTrue:  [ philosopher <-: denied ]
      ifFalse: [
        forks at: leftForkId  put: true. @*\label{ln:pick}*@
        forks at: rightForkId put: true. @*\label{ln:endpick}*@
        philosopher <-: eat ]
  ) @*\label{ln:endpickUpForks}*@
  public putDownForks: leftForkId = (
    | rightForkId |
    rightForkId := 1 + (numForks % leftForkId). @*\label{ln:nforks}*@
    forks at: leftForkId  put: false.
    forks at: rightForkId put: false.
  )
)\end{lstlisting}

In contrast to communication deadlocks, all variants of actor models can suffer from behavioral deadlocks.
One cause for such deadlocks are \emph{flexible interfaces}\cite{DeKoster:2016:YAT}, because when an actor limits the set of messages it accepts, the overall system can reach a state where actors mutually wait for messages being sent, without allowing any progress.
On the other hand, if an actor implements two or more interfaces, it could be that only one of them is deadlocked, allowing some progress with respect to interactions with other actors.

\subsubsection{Livelock.}
A program is in a livelock when an actor or a group of actors can make local progress, but the program is not able to make global progress.
For example, actors can change their state receiving and executing messages, but the overall execution of the program stalls and cannot be finished. 

An example for a livelock is given in \Cref{lst:barber}. It shows
the sleeping barber problem \cite{Dijkstra:1968} implemented in Newspeak\cite{Bracha:10:NS}. 
The waiting room, the barber, and the customers are implemented as actors. 
The concurrency issue in this example is caused by a \emph{fault} in \cref{ln:remove}. Instead of receiving the next customer from the collection of customers \code{waitingCustomers}, the barber always receives the same first customer. Both actors, room and barber are not blocked. The barber asks for the next customer to the room (\cref{ln:next}) and the room sends the customer to the barber to do the haircut (\cref{ln:enter}). But, as the customer that is sent is always the same, there is no global progress.

\begin{lstlisting}[caption={Livelock in a sleeping barber implementation. \Cref{ln:remove} reads always the same customer, but does not remove it from the list, preventing global progress.},label=lst:barber,float=tbhp] 
class WaitingRoomActor new: capacity barber: anActor = (
  (* ... *)
  public next = (
    waitingCustomers size > 0
     ifTrue: [
       | customer |
       customer := waitingCustomers first. @*\label{ln:remove}*@
       barber <-: enter: customer in: self ] @*\label{ln:enter}*@
     ifFalse: [
       barber <-: wait.
       barberAsleep := true ]
  )
)
class BarberActor new: resolver = (
  (* ... *)
  public enter: customer in: room = (
    customer <-: start.
    busyWait: (random next: avHaircutRate) + 10.
    customer <-: done.
    room <-: next @*\label{ln:next}*@
  )
)\end{lstlisting}

\subsection{Message Protocol Violations}

As shown in \Cref{tab:taxonomy}, thread-based programs commonly suffer from three sorts of low-level race conditions:  data races, bad interleavings (also know as high-level data race\cite{Artho:2003}, atomicity violation\cite{Abbaspour:2016}), and order violations.  
Actors, on the other hand, cannot suffer from those low-level race conditions since they have exclusive access to their state and messages are processed serially.
Nevertheless, all actor-based programs can have race conditions related to the order in which messages are processed.
We consider these race conditions to be at a \emph{high-level} to distinguish them from the low-level memory access race conditions that occur in thread-based programs.

High-level race conditions in actor based-programs can be observed when two or more actors exchange messages that are not consistent with the intended \emph{protocol} of the application. Therefore, we refer to them more specifically as \emph{message protocol violations}.
We identified three types of message protocol violations, which are described in the remainder of this subsection: \emph{message order violations}, \emph{bad message interleavings}, and \emph{memory inconsistencies}.

\subsubsection{Message order violation.}
A message order violation appears when the order in which two or more actors exchange messages is not consistent with the intended \emph{protocol} of the actor.
This includes messages that are received out of order or in unexpected interleavings. 
They are typically caused by actors only supporting a subset of all possible message sequences. 

Message order violations are common for instance in JavaScript.
In a contemporary browser, each script runs inside one single-threaded event-loop per page.
After the initial parsing and interpretation of \code{<script>} tags, the event-loop processes incoming events related to page lifecycle events, UI events, timer events, XRS responses, etc.
The order in which corresponding event handlers are executed is non-deterministic, \eg, because of user actions or I/O timing, which can give rise to an unexpected ordering of messages that is not handled correctly by the program. 
Listing \ref{lst:JS} extracted from\cite{Raychev:2013} shows an example of such a message order violation.
The \emph{fault} occurs in \cref{ln:f}, in this case because of an interleaving between the execution of the user action \code{onclick} and the HTML parsing.

\begin{lstlisting}[caption={Message order violation within a single event-loop in JavaScript (from \cite{Raychev:2013}). On \cref{ln:f}, the \code{onclick} event can be triggered by the user before the function \code{f} is parsed and made available, causing an error.},label=lst:JS, language=HTML,float=tbhp]
<html><body>
  <input type="button" id="b1" onclick="javascript:f()">	@*\label{ln:f}*@
   ... <!-- many elements -->
   <script>			@*\label{ln:s1}*@
   function f() {
     if (init)
       alert(y.g);
     else
       alert("not ready");
   }		
     var init = false, y = null;
   </script> @*\label{ln:s2}*@
     ...
   <script>
     y = { g: 42 };
    init = true;
   </script>
 </body></html>
 \end{lstlisting}

The code in \Cref{lst:JS} defines an input tag for a button in an HTML page (\cref{ln:f}), and two scripts: one declaring two variables (\code{init} and \code{y}) and the behavior of function \code{f} which is executed when the button is clicked (\cref{ln:s1}--\ref{ln:s2}), and a second script which updates the variables \code{init} and \code{y}.
 Since the parsing of the \code{input} tag and the execution of the scripts happen in different turns of the event-loop, a violation in the order of messages execution can occur. 
For instance, if the button is clicked before the first script runs, the function \code{f} is not yet declared,  causing the JavaScript interpreter to crash.

Note that message order violations in JavaScript only affect a \emph{single} actor, because a JavaScript program runs in a \emph{single} event-loop, which processes all types of events. 
General message order violations can also involve more than two actors.

\subsubsection{Bad message interleaving.}
We define a \emph{bad message interleaving} as the condition when a message is processed between two messages which are expected to be processed one after the other, causing some misbehavior of the application or even a crash. 

In the original actor model, when an actor sends a message to a recipient actor, the message is placed in a mailbox and is guaranteed to be eventually delivered by the actor system. 
All messages are thus expected to be delivered in the order in which the sender actor sent them.
However, there are two sources of bad interleavings.
First, messages from different senders may be interleaved in between messages from one sender.
In other words, even if the actor model enforces that messages from a sender actor are received in a FIFO order, messages from different sender actors may occur between them.
The second source of bad interleavings of messages occurs in variants of the actor model which do not guarantee in-order delivery of the messages.
This can be found in actor models used to build distributed systems, like Scala or ActorFoundry \cite{Lauterburg:2009} in which communication between actors is not enforced to work in a FIFO manner.

\begin{lstlisting}[caption={Bad message interleaving example in ActorFoundry (from \cite{Lauterburg:2009}). The \code{Server} actor can interleave the messages \code{set} and \code{get} send by the \code{Client}. If that is the case \code{v1} will a value that differs from \code{v2}.}, label={lst:bad-interleaving}, language=Java,float=tbhp]
class Server extends Actor {
  int value = 0;
  @message void set(int v) { value = v; }
  @message int  get()      { return value; }
}
class Client extends Actor {
  ActorName server;
  Client(ActorName s) { server = s; }
  @message void start() {
    send(server, "set", 1); @*\label{ln:client-send}*@
    int v1 = call(server, "get"); @*\label{ln:client-get}*@
    int v2 = call(server, "get");
    assert v1 == v2; @*\label{ln:assertion}*@
  }
}\end{lstlisting}

\Cref{lst:bad-interleaving} shows an example of bad message interleavings in ActorFoundry (extracted from \cite{Lauterburg:2009}).
The listing shows an example of bad message interleaving in a network communication between two actors, \code{Server} and \code{Client}.
In \cref{ln:client-send}, the \code{Client} sends an asynchronous message to the \code{Server} to store the value 1. In \cref{ln:client-get}, the \code{Client} does a \code{call}, which waits for a result, to retrieve the value from the \code{Server}. 
The \emph{fault} is triggered by \cref{ln:assertion}, because it can happen that the \code{Server} processes the \code{set} message between the two  \code{get} messages. Consequently, the values of \code{v1} and \code{v2} will be inconsistent.

Note that in the context of JavaScript, bad message interleavings can also occur within a single event-loop if programs can receive notifications for external events, \eg events from the network, from timers or from sensors.
Such issues have been previously reported by \cite{Hong:2014}.

\subsubsection{Memory inconsistency.}

A memory inconsistency is a condition in which different actors have inconsistent views of shared resources. 
This can be caused because the effects of the turn that modifies a \emph{conceptually shared resource} may not be visible to other actors which also alter the same resource.
Previous research on Erlang has collected such kinds of problems \cite{Huch:1999, Hughes:2011, DOsualdo2013}. 

\Cref{lst:memory} shows a modified fragment of an Erlang program used by D'Osualdo et al. \cite{DOsualdo2013} to verify the property of mutual exclusion in actors.
The program (originally introduced by Huch \cite{Huch:1999}) spawns one database process and several client processes. 
The purpose of the program is to save information in a database, which acts as a conceptually shared resource by different client actors.
The database consists of a map of key-value tuples. %
When a client process sends an \code{allocate} message to the database, the database checks if the key exists already (\cref{ln:look}). 
If the value does not exist (\cref{ln:key}) then it is saved. 
The \code{free} message in the client computes the value to be saved (\cref{ln:free}) and then the client process sends the tuple to the database.
If a second process does lookup before the first value is saved, the \code{lookup} function will fail due to the key not having been inserted yet.
The \emph{fault} occurs in \cref{ln:value}, when the database process receives the key and value to be stored.
Another client that has a different value with the same key can save it. 
Thus, the value sent by the first process will be overwritten by the value of another client process.
To fix this error, the message pattern should be declared inside a \code{receive} statement after \cref{ln:free} to save the value sent by the client and avoid other processes making a lookup. 

\begin{lstlisting}[caption={Memory inconsistency example in Erlang (based on \cite{Huch:1999, DOsualdo2013}). Line \ref{ln:value} shows a message pattern that allows different processes to store different values for the same key.}, label={lst:memory}, language=Erlang,float=tbhp]
main() ->
    DB = spawn(fun()->dataBase(#{})end),
    spawnmany(fun()->client(DB) end).
    
dataBase(M) ->
   receive
       {allocate,Key,P} ->
           case lookup(Key,M) of  @*\label{ln:look}*@
               fail ->
                   P!free,        @*\label{ln:free}*@
                   dataBase(M);                
               succ ->
                   P!allocated,
                   dataBase(M)
           end;
       {lookup,Key,P} ->
           P!lookup(Key,M),
           dataBase(M);
       {value,Key,V} ->           @*\label{ln:value}*@
          dataBase(maps:put(Key,V, M))
   end.

lookup(K,M) -> 
   case maps:find(K,M) of
       error -> fail;             @*\label{ln:key}*@
       _V     -> succ
   end.
	
\end{lstlisting}

\subsection{Comparison with Existing Terminology in Actor Literature}

As pointed out in the introduction, the goal of establishing a taxonomy is to provide a common vocabulary for concurrency bugs in actor-based programs.
In what follows we relate our terminology to the one presented in other efforts tackling concurrency bugs for actor-based programs.

Bad message interleavings have been denoted as \emph{ordering problems} by Lauterburg et al. \cite{Lauterburg:2009} and Long et al. \cite{Long:2016} and as \emph{atomicity violation} by Zheng et al. \cite{Zheng:2011} and Hong et al. \cite{Hong:2014}. 
We consider ordering problems to be too coarse-grained terminology.
We decided to use the term bad message interleaving to avoid confusion with atomicity violations in thread-based concurrent programs due to low-level memory accesses errors.

Message order violations have been collected under many different names in literature: \emph{data races} by Petrov et al. \cite{Petrov:2012}, \emph{harmful races} by Raychev et al. \cite{Raychev:2013}, \emph{order violations} by Hong et al. \cite{Hong:2014}, and \emph{message ordering bugs} by Tasharofi et al. \cite{Tasharofi:2013}.
We consider message order violations to be a descriptive name while avoiding confusion with low-level data races present in thread-based programs.

Memory inconsistency problems have been denoted as \emph{race conditions} by Hughes and Bolinder \cite{Hughes:2011}. D'Osualdo \cite{DOsualdo2013} tackled this problem by proving a correctness property referred to as ``mutual exclusion''.

In literature, the term \emph{orphan messages} \cite{Colaco1997} refers to messages that an actor sends but that the receiver actor(s) will never handle.
Rather than a kind of concurrency bug, we consider orphan messages as an observable property of an actor system which may be a symptom of a concurrency bug like communication deadlocks or message ordering violations.
We use this terminology in the next section when we classify concurrency bugs reported in literature with our taxonomy.
Orphan messages can for example be present in actor languages that allow flexible interfaces such as Erlang, the Scala Actors framework and the Akka library \cite{DeKoster:2016:YAT}. An actor may change the set of messages it accepts after another actor has already sent a message which can only be received by an interface which is no longer supported.

\section{Concurrency Bugs in Actor-based Programs} 
\label{sec:sota:bugs}

In this section, we review various concurrency bugs reported in literature, and classify them according to the taxonomy introduced in \Cref{sec:bugs}.
The goal is twofold:
(1) to classify concurrency bugs collected in prior research in the bug categories according to our taxonomy and (2) to identify bug patterns and observable behaviors that appear in programs exhibiting a particular concurrency bug. 
The latter is useful to design mechanisms for testing, verification, static analysis, or debugging of such concurrency issues.

\Cref{tab:bugs} shows the catalog of analyzed concurrency bugs collected from literature. 
In the first column we categorized these bugs according to the taxonomy presented in \Cref{tab:taxonomy}. 
For each bug scenario we describe the bug pattern as a generalized description
of the fault by identifying the actions that trigger the error.
In the remainder, we highlight the identified bug patterns in italic.
We also describe the observable behavior of the program that has the concurrency issue, \ie the failure.

\subsection{Lack of Progress Issues}

To the best of our knowledge, the literature reports on communication deadlocks mostly in the context of Erlang programs.
Bug-4 in \Cref{tab:bugs} is an example of a communication deadlock collected by Christakis and Sagonas \cite{Christakis:2011}, which corresponds to the example depicted in \Cref{lst:pingpong}.
Christakis and Sagonas \cite{Christakis:2011} distinguish two causes for communication deadlocks in Erlang programs:
\begin{itemize}
\item \emph{receive-statement with no messages} \ie empty mailbox, 
\item \emph{receive with the wrong kind} \ie the messages of the mailbox are different to the ones expected by the receive statement.
\end{itemize} 
We classify these conditions as bug patterns for orphan messages, which can lead to communication deadlocks in Erlang. 

Christakis and Sagonas \cite{Christakis:2011b} mention also other conditions that can cause mailbox overflows or potentially indicate logical errors. 
Such conditions include \emph{no matching receive}, \ie the process does not have any \code{receive} clause matching a message in its mailbox, or
\emph{receive-statement with unnecessary patterns}, \ie the \code{receive} statement contains patterns that are never used. 

Bug-9 is similar in kind to bug-4. Bug-9 was identified by Gotovos et al. \cite{Gotovos:2011} when implementing a test program in Erlang which has a server process that receives and replies to messages inside a loop. 
The server process blocks indefinitely because it waits for a message that is never sent.
They also identify it as problematic, \emph{when a message is sent to an already finished process}, which is exhibited by bug-10. 
This can happen due to two possible situations.
First, if a client process sends a message to an already finished server process, the client process will throw an exception.
Second, if the server process exits without replying after the message was received, the client process will block waiting for a reply that is never sent.
We categorize bug-4, bug-9, and bug-10 as communication deadlocks and the observable behaviors as orphan messages.

D'Osualdo et al. \cite{DOsualdo2013} identified three other bug patterns leading to abnormal process termination in Erlang programs, which might cause deadlocks:
\emph{sending a message to a non-pid value}, 
\emph{applying a function with the wrong arity} and 
\emph{spawning a non-functional value}. 
These bug patterns could result in a communication deadlock or in a message order violation if the termination notification is not handled correctly.

Aronis and Sagonas \cite{Aronis:2017} studied built-ins operations that can cause races in Erlang programs.
Because the studied built-ins can access memory that is shared by processes, races can be observed in form of different outputs.
Their classification on observable interferences of Erlang/OTP built-ins can help to diagnose communication deadlocks, message order violations, and memory inconsistencies.

\subsection{Message Protocol Violations}

\subsubsection{Message order violation.}

In Erlang, updating certain resources such as the global name registry requires 
careful coordination to avoid concurrency issues. For example, we categorize
bug-1 as a message order violation, which as a result makes a race on the
global process registry visible\cite{Christakis:2010}.
The bug is caused because two processes try to register processes for the
same global name more than once, which is done with non-atomic operations.
For correctness, these processes would need to coordinate with each other.

Bug-11 reported by Christakis et al. \cite{Christakis:2013} is another example of a message order violation exhibited when a \emph{spawned process terminates before the parent process registers its process id}. 
The application expects the parent process to register the id of the spawned process before the spawned process is finalized, but as the execution of \code{spawn} and \code{register} functions are not atomic, an unexpected termination can cause a message order violation.

Zheng et al. \cite{Zheng:2011} studied concurrency issues that can appear in JavaScript programs. In their example, which corresponds to bug-14, two events are executed  but the application cannot return the responses in time, \eg \emph{the second message is executed with the value of the first message}. 
They argue that the cause of this issue can be the network latency and the delay in managing the responses by the JavaScript engine.  If the events operate on the same data, it can lead to inconsistencies \eg deleting an object of a previous event.
We consider this kind of race as a message order violation, because the order of the messages is not consistent with the protocol of the web application.

In the context of JavaScript, Petrov et al. \cite{Petrov:2012} identified 4 different message order violations.
An \emph{interleaving between the execution of a script and the event for rendering an input text box} is shown in bug-17, which can lead to inconsistencies when saving the text a user entered.
Also problematic is the potential \emph{interleaving of creating an HTML element and
executing a script that uses the element} shown in bug-18. If the HTML element has not yet been created, it will cause an exception.
Moreover, bug-19 corresponds to the scenario where \emph{executing a function can race with is definition}.
This can happen when the function is invoked first because the HTML loads faster, and the script where it is declared is only loaded later.
For example in bug-20, \emph{the \code{onload} event of an HTML element is triggered before the code is loaded}, which causes the event handler to never run correctly.

Raychev et al. \cite{Raychev:2013} detected similar race conditions to the one of Petrov et al. \cite{Petrov:2012}, which we categorize as message order violations. Their bug example is depicted in \Cref{lst:JS} and corresponds to bug-16.
Hong et al. \cite{Hong:2014} also collected message ordering violations in three different existing websites. One of its examples shows a scenario where \emph{a user input invokes a function before it is defined}. This last example is detailed in bug-23. 
From all these collected bugs, we conclude that a common issue in JavaScript programs is the \emph{bad interleaving of two events 
in an unexpected order}. %
 
Tasharofi et al. \cite{Tasharofi:2013} identified twelve bugs in five Scala projects using the Akka actor library, which we categorize as message ordering problems.
Bug-13 gives details of one of these bugs.
The study found two bug patterns in Scala and Akka programs that can cause concurrency bugs in actors.
First, \emph{when changing the order of two receives in a single actor (consecutive or not)}, which can provoke a message order violation.
Second, \emph{when an actor sends a message to another actor which does not have the suitable receive for that message}. 
This last issue corresponds to an orphan message, and can also lead to other misbehaviors such as communication deadlocks.

\subsubsection{Bad message interleaving.}

Bug-12 corresponds to the example of bad message interleaving collected by Lauterburg et al. \cite{Lauterburg:2009} which was shown in \Cref{lst:bad-interleaving}. The bug pattern occurs when \emph{an actor executes a third message between two consecutive messages due to the actor model implementation being not FIFO}.

Zheng et al. \cite{Zheng:2011} also identified bad message interleavings such as the one exhibited in bug-15. The bug pattern corresponds to the \emph{use of a variable not initialized by other methods before it was defined}.
This delay of receiving a response can be caused by a busy network and leads to an exception in the application.
Hong et al. \cite{Hong:2014} also observed bad message interleavings in JavaScript programs.
Bug-21 shows a pattern in which a variable is undefined because \emph{after a user has uploaded a file to a workspace, the user changes the workspace before the file has been completely uploaded}.
In the case of bug-22, a variable is \code{null} because an \emph{event handler updates the DOM between two inputs events that manipulate the same DOM element}.

\subsubsection{Memory inconsistency.}
To the best of our knowledge, memory inconsistency issues have only been reported in the context of Erlang programs.
Christakis and Sagonas \cite{Christakis:2010} shows an example of high-level races between processes using the Erlang Term Storage in bug-2. In this case the error is due to \emph{inserting and lookup in tables that have public access}, thus it is possible that two or more processes try to read and write from them simultaneously.
A second example detailed in bug-3, shows a similar issue that can happen when accessing tables of the Mnesia database. The cause is due to the \emph{use of reading and writing operations that can cause race conditions}.
We categorize both issues as memory inconsistency problems.

Hughes and Bolinder \cite{Hughes:2011}  detected four bugs corresponding to memory inconsistencies in \emph{dets}, the disk storage back end used in the Erlang database Mnesia. 
Bug-5 refers to \emph{insert operations that run in parallel} instead of being queued in a single queue. They can cause inconsistent return values or even exceptions.
The observable behavior of bug-6  corresponds to an inconsistency of visualizing the dets content. This issue can occur when \emph{reopening a file that is already open and executing insert and get\_contents operations in parallel}. 
Bug-7 and bug-8 are caused due to failure on integrity checks. Of the four bugs that were found, these two are the ones that can occur with the least probability. 
Bug-7 is reproduced only in one specific scenario when \emph{running three processes in parallel}, and bug-8 can occur only in those languages implementations that \emph{can keep new and old versions of the server state}. 

Huch \cite{Huch:1999} and D'Osualdo et al. \cite{DOsualdo2013} conducted studies to verify mutual exclusion in Erlang programs. \Cref{lst:memory} shows an example. 
The bug pattern identified corresponds to the \emph{wrong definition of the behavior of the actor}, and the observable property is that two actors can store different values for the same key which leads to inconsistencies, \ie the actors can share the same resource.

\subsection{Actor Variants and Possible Bugs}
Based on our review of concurrency bugs above, we summarize which concurrency bugs can occur for each variant of the actor model.
Furthermore, we identify the patterns that can cause a concurrency bug and the behavior that can be observed in the programs that have these bugs.
In languages that implement the \emph{process} actor model, e.g. Erlang and Scala, programs can exhibit communication deadlocks because the actor implementation provides blocking operations. 
A common observable behavior of this concurrency bug are the orphan messages. 
This means an actor with this issue is blocked, i.e. the process is in a waiting state.
These languages can also suffer from message order violations and memory inconsistencies.
For message order violations possible bug patterns are the delays in managing responses, or the unsupported interleaving of messages i.e. the actor protocol does not correspond to the executed message interleavings.
These can result in a program crash or inconsistent computational results.
Memory inconsistencies are typically caused by a wrong message order when accessing shared resources.
Languages such as AmbientTalk or JavaScript that use the \emph{communicating event-loop} model do not provide blocking primitives, and thus, do not suffer from communication deadlocks.
However, other lack of progress issues such as behavioral deadlocks and livelocks can occur.
Bug patterns for a behavioral deadlock or a livelock are typically mistakes in the sequential code of the actor, or a message that was sent to the wrong actor at the wrong time.
The resulting observable behavior can be a wrong program output in which one or more actors do not progress with their computation.
Behavioral deadlocks are possible in all variants of actor models.
They are one of the most difficult bugs to identify,
because actors are not blocked, but do not make any progress.
Livelocks are similarly hard to diagnose as behavioral deadlocks.

Similarly to the process actor variant, event-loop based programs can suffer from message order violations and bad message interleavings.
Generally, message order violations, bad message interleaving, and memory inconsistencies are race conditions that can happen in all actor-based programs including in programs using the class or \emph{active object} actor model variants.

\section{Advanced Development Techniques}

This section surveys the current state of the art of techniques that support
the development of actor-based programs. The goal is to identify the relevant
\emph{subfields of study} and \emph{problems} in the literature. Furthermore, for each
of these techniques we analyzed based on the literature how they relate to the
bug categories of our taxonomy to identify open issues.

Specifically, we survey techniques for static analysis, testing tools, debuggers, and visualization.
\Cref{tab:techniques} gives an overview of the categories of bugs that static analysis and testing techniques address. It leaves out debugging and visualization techniques, since they are typically not geared towards a specific
set of bugs.

\subsection{Static Analysis}

The static analysis approaches surveyed in this section include all approaches
that identify concurrency issues without executing a program. This includes
approaches based on typing, abstract interpretation, symbolic execution, and
model checking. The following descriptions are organized by the category of concurrency bugs these approaches address.

\subsubsection{Lack of progress issues.}
In the field of actor languages, Erlang has been subject to extensive studies.
Dialyzer is a static analysis tool that uses type inference in addition to type
annotations to analyze Erlang code\cite{Sagonas:2005}. 
The static analysis uses information on control flow and data flow to identify problematic usage of Erlang built-in functions that can cause concurrency issues.
Dialyzer also has support for
detecting message order violations as well as memory
inconsistencies\cite{Sagonas:2010,Christakis:2010}.
Christakis and Sagonas \cite{Christakis:2011} extended Dialyzer to also detect communication
deadlocks in Erlang using a technique based on \emph{communication graphs}.

Another branch of work uses type systems to prevent concurrency issues. For
actor languages, this includes for instance the work of Cola\c{c}o et al.\cite{Colaco1997}.
Based on a type system for a primitive actor calculus, they can prevent many
situations in which messages would be received but never processed, \ie,
so-called orphan messages. However, static analysis cannot detect all
possible orphan messages. Therefore, the approach relies on \emph{dynamic type checks} to detect the remaining cases.
Similar work was done for Erlang, where orphan messages are also
detected based on a type system\cite{Dagnat:2002}.

Dam and Fredlund \cite{Dam:1998:VOD} proposed an approach using  static analysis to verify properties such as the boundedness of mailboxes. 
The verification of this property can avoid the presence of orphan messages in a program.
Their technique applies \emph{local model checking in combination with temporal logic and extensions to the $\mu$-calculus} for basic Erlang systems.

Similarly, Stiévenart et al. \cite{Stievenart:2017} used \emph{abstract interpretation} techniques to statically verify the absence of errors in actor-based programs and upper bounds of actor mailboxes.
As mentioned before the verification of mailbox bounds can avoid the presence of orphan messages. 
The proposed technique is based on different mailbox abstractions which allows to preserve the order and multiplicity of the messages.
Thus, this verification technique can be useful to avoid message order violations.

\subsubsection{Message protocol violation.}
D'Osualdo et al. \cite{DOsualdo2013} also worked on Erlang and used static analysis and
\emph{infinite-state model checking}. Their goal is to check specific properties for
programs that are expressed with annotations in the code. With this approach,
they are able to verify for instance correct mutual exclusion semantics modeled
with messages. However, their current approach cannot model arbitrary message
order violations, because the used analysis abstracts too coarsely from
messages.

Garoche et al. \cite{Garoche:2006} verify safety properties statically for an actor calculus
by using \emph{abstract interpretation}. Their work focuses on orphan messages and
specific message order violations. Their technique is especially suited 
for detecting unreadable behavior, detecting unboundedness of resources,
and determining whether linearity constraints hold.

Zheng et al. \cite{Zheng:2011} developed a static analysis for JavaScript relying
on \emph{call graphs and points-to sets}. The analysis detects bad message
interleavings and message order violations. With the properties of
JavaScript, one can consider this analysis as a special case for actor systems
where only a single actor is analyzed with respect to its
reaction to incoming messages.
WebRacer\cite{Petrov:2012} is a tool that uses a \emph{memory access model and a
notion of happens-before relations} for detecting races at the level of the DOM
tree nodes. The detected bugs correspond to bad message interleavings and message
order violations in our taxonomy.
EventRacer\cite{Raychev:2013} is another tool that aims at finding bad message
interleavings or message order violations in JavaScript applications. In
this case the authors proposed a race detection algorithm based on \emph{vector
clocks}.

\subsection{Testing Tools}

This section describes work on testing actor based-programs to identify concurrency bugs. Some of the approaches are based on recording the interleaving of messages, the usage of state model checkers, and techniques to analyze message schedules.

\subsubsection{Lack of progress issues.}
Sen and Agha \cite{Sen:2006} present an approach to detect communication deadlocks in a language closely
related to actor semantics. They use a \emph{concolic testing} approach that combines
symbolic execution for input data generation with concrete execution to
determine branch coverage. The key aspect of their technique is to minimize the
number of execution paths that need to be explored while maintaining full
coverage.

Concuerror\cite{Christakis:2013} is a systematic testing tool for Erlang
that can detect abnormal process termination as well as blocked processes, which might indicate a communication deadlock. To
identify these issues, Concuerror \emph{records process interleavings} for test
executions and implements a stateless search strategy to explore all
interleavings.

\subsubsection{Message protocol violation.}
Claessen et al. \cite{Claessen:2009:FRC} use a \emph{test-case-generation approach} based on
QuickCheck in combination with a custom user-level scheduler to identify race
conditions. The focus is specifically on bad message interleavings and process
termination issues. To make their approach intuitive for developers, they visualize problematic traces.
Hughes and Bolinder \cite{Hughes:2011} use the same approach and apply it to a key component of
the Mnesia database for Erlang. They demonstrate that the system is able to
find race conditions at the message level that can occur when
interacting with the shared memory primitives used by Mnesia.

Basset\cite{Lauterburg:2009, Lauterburg:2010} is an automated testing tool
based on Java PathFinder, a \emph{state model checker}, that can discover bad message interleavings in Scala and ActorFoundry programs. 
\cite{Tasharof:i2012} improve Basset with a technique to reduce schedules to
be explored, which improves the performance of Basset. Their key insight is to
exploit the transitivity of message send dependencies to prune the search space
for relevant execution schedules.
For the Scala-Akka programs there is another testing tool called Bita, which can also detect message order violations. Their proposal is based on a technique called schedule coverage, which analyzes the order of the receive events of an actor\cite{Tasharofi:2013}.

The Setac framework\cite{TasharofiSetac:2011} for the Scala Actors framework enables testing for race conditions on
actor messages, specifically message order violations. A test case defines \emph{constraints on schedules and assertions} to
be verified, while the framework identifies and executes all relevant schedules
on the granularity of message processing.
The Akka actor framework for Scala also provides a test framework called
\citeurl{TestKit.}{Akka.io: Testing Actor Systems}{Lightbend Inc.}{8 February 2017}{http://doc.akka.io/docs/akka/current/scala/testing.html}
However, it does not seem to provide any sophisticated automatic testing capabilities, which seems to indicate that the current techniques might not yet
be ready for adoption in industry.

Cassar and Francalanza \cite{Cassar:2014} investigate how to minimize the overhead of instrumentation
to detect race conditions. Instead of relying exclusively on synchronous
instrumentation, they use \emph{asynchronous monitoring} in combination with a logic
to express correctness constraints on the resulting event traces.

Hong et al. \cite{Hong:2014} proposed a JavaScript testing framework called WAVE for the
same classes of issues mentioned by \cite{Petrov:2012} and \cite{Raychev:2013}. The framework \emph{generates test cases based on operation sequences}. 
In case of a concurrency bug, they can observe different results for the generated test cases.

\begin{table}
\centering
\begin{tabular}{l c c c c c c c}

\hline
 & Communi.  & Behav.   & Live- & Message Or. & Bad Msg. & Mem.   \\
 & Deadlock  & Deadlock & Lock  & Violation   & Inter.   & Incon. \\
\hline
\emph{Static Analysis} & \\

Christakis and Sagonas \cite{Christakis:2011}              & X  &  \\
Christakis and Sagonas \cite{Christakis:2010}              &    &   &   & X &   & X    \\
Cola\c{c}o et al. \cite{Colaco1997}                   & p  &   &   &   &   &    \\
Dagnat and Pantel \cite{Dagnat:2002}             & p  &   &   &   &   &    \\
Dam and Fredlund \cite{Dam:1998:VOD}				 & p  &   &   &   &   &    \\
Stiévenart et al. \cite{Stievenart:2017}				 & p  &   &   & p  &   &    \\
D'Osualdo \cite{DOsualdo2013}                 & p  &   &   & p &  & p  \\
Garoche et al. \cite{Garoche:2006}                 & p  &   &   & p &   &    \\
Zheng et al. \cite{Zheng:2011}                   &    &   &   & p & p &    \\
Petrov et al. \cite{Petrov:2012}        			 &   &   &   & X & X &      \\
Raychev et al. \cite{Raychev:2013}       			 &   &   &   & X &  &      \\
\hline
\emph{Testing Tools} & \\

Sen and Agha \cite{Sen:2006}           & X &   &   &   &   &      \\
Claessen et al. \cite{Claessen:2009:FRC}  &   &   &   &   & X &      \\
Christakis et al. \cite{Christakis:2013}    & X &   &   &   &   &      \\
Lauterburg et al. \cite{Lauterburg:2010}    &   &   &   &   & X &      \\
Tasharofi et al. \cite{Tasharofi:2013}     &   &   &   &   & X &      \\
Tasharofi et al. \cite{TasharofiSetac:2011} &   &   &   & p & p &   \\
Tasharofi et al. \cite{Tasharof:i2012}     &   &   &   & p & X &   \\
Hughes and Bolinder \cite{Hughes:2011}        &   &   &   & p &   & X  \\
Hong et al. \cite{Hong:2014}          &   &   &   & X & X &      \\
Cassar and Francalanza \cite{Cassar:2014}        &   &   &   & p & p & p  \\
\hline

\end{tabular}
\caption{Overview of the bug categories addressed in literature.
A `p' indicates that a bug category is addressed only partially.
Typically, the approaches are limited by, for instance, a too coarse abstraction
or a description language not expressive enough to capture all bugs in a category.}
\label{tab:techniques}
\end{table}

\subsection{Debuggers}
\label{sec:debuggers}
This section reviews the main features provided by current debuggers for actor-based systems. 
It includes techniques for both online and postmortem debugging.

Causeway\cite{Stanley:2009} is a postmortem debugger for distributed communicating event-loop programs in E\cite{Miller:2005}.
It focuses on displaying the \emph{causal relation of messages} to enable developers to determine the cause of a bug.
Causality is modeled as the partial order of events based on Lamport's happened-before relationship\cite{Lamport:1978}. 
We consider that this approach can be useful for detecting message protocol violations.
REME-D\cite{Gonzalez:2014} is an online debugger for distributed communicating event-loop programs written in AmbientTalk\cite{Vancutsem:2007}. 
REME-D provides message-oriented debugging techniques such as the \emph{state inspection}, in which the developer can inspect an actor's mailbox and objects, while the actor is suspended. It also supports a catalog of breakpoints, which can be set on asynchronous and future-type messages sent between actors.
Like Causeway, REME-D allows inspecting the history of messages that were sent and received when an actor is suspended, also known as \emph{causal link browsing}\cite{Gonzalez:2014}. 
Therefore, we consider debugging techniques provided in REME-D to be helpful for detecting message order violations. Also the technique of inspecting the state of the actor can facilitate debugging any lack of progress issues such as behavioral deadlocks and livelocks.
Kómpos\cite{Marr:2017} is an online debugger for SOMns.
For debugging actor-based programs, Kómpos provides a wide set of \emph{message-oriented breakpoints and stepping operations}.
For example, Kómpos breakpoints allow developers to inspect the program state before a message is sent or after the message is received, but before it is processed on the receiver side.
Moreover, is possible to pause the program execution before a promise is resolved with a value or before the first statement of a callback to that promise is executed, i.e. once the promise has been resolved.
Breakpoints to pause on the first and last statement of methods activated by an asynchronous message sent can be also set. 
Stepping operations can be triggered from the mentioned breakpoint locations.
Furthermore, one can continue the actor's execution and pause in the next turn or pause before the execution of the first statement of a callback registered to a promise.
This set of debugging operations gives more flexible tools to actor developers to deal with lack of progress issues such as behavioral deadlocks and livelocks.
In addition, a specific actor visualization is offered that shows actor turns and messages sends. 
This can be useful when trying to identify the root cause of a message protocol violation.

In the context of JavaScript, the Chrome DevTools online debugger supports \citeurl{Web Workers,}{Web Workers}{W3C}{14 February 2017}{https://www.w3.org/TR/workers/}
which are actors that communicate with the main actor through message passing.
The Chrome debugger allows pausing \emph{workers}. In the case of \emph{shared workers} it also provides mechanisms to inspect, terminate, and set breakpoints.\footnote{\url{http://blog.chromium.org/2012/04/debugging-web-workers-with-chrome.html}}
For debugging messages and promises on the event-loop, Chrome also supports \emph{asynchronous stack traces}.
This means, it shows the stack at the point a callback was scheduled on the event-loop. Since this works transitively, it allows inferring the point and context of how a callback got executed.
We consider that stack information could help finding both message order violation and lack of progress issues. 

Erlang also has an \citeurl{online debugger}{Debugger}{Ericsson AB}{14 February 2017}{http://erlang.org/doc/apps/debugger/debugger_chapter.html} that supports \emph{line, conditional, and function breakpoints}. The Erlang processes can be inspected from a list and
for each process a view with its current state as well as its current location
in the code can be opened, which allows one to inspect and interact with each process independently.
It also supports stepping through processes and inspecting their state.
We consider that process inspection information could help finding both message protocol violations and lack of progress issues. 
The ScalaIDE also includes facilities for debugging of \citeurl{actor-based programs.}{Asynchronous Debugger}{ScalaIDE}{14 February 2017}{http://scala-ide.org/docs/current-user-doc/features/async-debugger/index.html} It is a classic online debugger with support for stepping, line and conditional breakpoints.
Furthermore, one can follow a message send and \emph{stop in the receiving actor}.
Additionally, the debugger supports asynchronous stack traces similar to Chrome\cite{Dragos:2013}.
We consider these techniques useful for debugging message protocol violations. They can also be used to identify behavioral deadlocks and livelocks when inspecting the state of the receiving actor.

The recently proposed Actoverse debugger\cite{Shibanai:2017} enables \emph{reverse debugging} of Akka programs written in Scala.
It uses snapshots of the state of actors to enable back-in-time debugging in a postmortem mode.
Furthermore, Actoverse provides message-oriented breakpoints and a message timeline that visualizes the messages exchanged by actors similar to a sequence diagram. 
The authors aim to ease finding the cause of message protocol violations in Akka programs.

\subsection{Visualization}

This section discusses mechanisms and approaches to visualize actor-based systems for debugging. Some of the techniques represent actor communication flow with petri nets. Other techniques detail an actor's state, its mailbox, and the traces of causal messages that are sent and received.

Miriyala et al. \cite{Miriyala:1992} proposed the use of \emph{predicate transition nets} for visualizing actors execution.
Based on the classic model of actors the proposal focus on the representation of the actor behavior and sent messages.
The activation of each transition in the petri net corresponds to a behavior execution.
The authors emphasize that the order of net transitions should be represented in the same order as the execution of messages of the actor system. 
The main idea is that the user interacts with a visual editor for building the execution of an actor system in the petri net. 

Coscas et al. \cite{Coscas:1995} present a similar approach in which the predicate transition nets are used to \emph{simulate actors execution in a step by step mode}. When a user fires a specific transition he or she only observes a small part of whole net. 
The approach also verifies messages that do not match with the ones expected by the actor, \ie messages that do not match the actor's interface.

The Causeway debugger also visualizes the program's execution based on views for \emph{process order, message order, stack and source code view}\cite{Stanley:2009}.
The \emph{process order} view shows all messages executed for each actor in chronological order, \eg a parent item with asynchronous message sends.
The \emph{message order} view shows the causal messages for a message sent, \ie other messages that have been executed before the message was sent and provoked the send of the message we want to debug. In this view it is also possible to distinguish processes by color, which helps users to visualize when a message flow (known as activation order) corresponds to a different process. 
The \emph{stack view} shows a partial causality of messages. It is considered partial because the call chain shown in the stack only visualizes the messages that have been executed, it does not show the other possible messages that can cause the invocation of a message (known as happened-before relation).
The \emph{source code view} shows the code where the message was sent in the code. Thanks to the synchronization achieved between all the views it is possible to transit through the messages related to the execution of the actor's behavior that led to the bug.

Gonzalez Boix et al. \cite{Gonzalez:2014} show the actor state in their REME-D debugger. The \emph{actor view} shows messages that are going to be executed in the actor's mailbox. At the same time it is also shown the state of the actor and its objects. 
This view is useful for the user in order to be able to interact with the objects and messages of the actor that is inspected. One of the main advantages of this online debugger is the possibility of pausing and resuming the actor's execution.

Recently, Beschastnikh et al. \cite{Beschastnikh:2016} developed ShiViz, a visualization tool where developers can visualize logs of distributed applications. The mechanism is based on \emph{representing happens-before relationships of messages} through interactive time-space diagrams. 
The tool also offers search fields by which messages can be searched in the diagram using keywords. Additionally, it is possible to \emph{find ordering patterns}, which could help to identifying wrong behaviors in an execution. 

\section{Conclusion and Future Work}
To enable research on debugging support for actor-based programs, we proposed a taxonomy of concurrency bugs for actor-based programs.
Although the actor model avoids data races and deadlocks by design, it is still possible to have lack of progress issues and message-level race conditions in actor-based programs.

Our literature review shows that actor-based programs exhibit a range of different issues depending on the specific actor model variant.
In languages like Erlang and Scala programs can suffer from communication deadlocks because the actor implementation uses blocking operations. 
In languages that implement the event-loop concurrency model this issue cannot occur. However, they can suffer from other lack of progress issues such as behavioral deadlocks and livelocks. 
Behavioral deadlocks and livelocks are really hard to identify because actors are not blocked, but still do not make any progress. 
Both lack of progress issues can be seen in all variants of the actor model.
Message order violations, bad message interleaving and memory inconsistencies are race conditions that can happen also in programs that implement any of the variants of the actor model.

Most work on identifying concurrency bugs is done in the fields of static analysis and testing.
Current techniques are effective for some specific cases, but often they are not yet general and do not necessarily scale to the complexity of modern systems.
Debugging support for actor languages currently provides features such as message-oriented breakpoints, inspecting the history of messages together with recording their casual relations, and support for asynchronous stack traces. 
However, better tools are needed to identify the cause of complex concurrency bugs. 

\subsubsection{Future work.}
For future work, there seems to be an opportunity for debuggers that combine strategies such as recording the causality of messages with message-oriented breakpoints and rich stepping.
Today, few debuggers support a full set of breakpoints that for example, allows one to debug messages stepping on the sender and on the receiver side. From the debuggers investigated in \Cref{sec:debuggers} only Kómpos allows us to set breakpoints on promises to inspect the computed value before it is used to resolve the promise.
We argue that the implementation of flexible breakpoints that adjust to the needs of actor-based programs is needed.
For instance, a breakpoint set on the sender side of the message will suspend an actor's execution before the message is sent. 
This can be useful when debugging lack of progress issues such as livelocks and behavioral deadlocks
because the developer will be able to see whether the message has the correct values.
Ideally, a debugger does not only allow us to inspect the turn flow, but to also combine the message stepping with the possibility of seeing the sequential operations that the actor executes inside of a turn. This gives developers better ways to identify the root cause of a bug.

Currently, only few debuggers allow developers to track the causality of messages. However, we consider this an important debugging technique. 
Recording the causal relationships of messages can help diagnosing, \eg, message protocol violations.
Back-in-time debugging techniques could be of great benefit for this. They are often used for postmortem debugging, because they allow developers to identify message order violations. 

Moreover, visualization techniques could be explored to give developers a better understanding of the debugging information.
To offer better visual support for actor systems, a combination of information about the actor's state and its objects, visualizing the order of execution of messages and including the happens-before relation between them, together with stack information 
should give the user better comprehension about the program that is debugged.
Nevertheless, further research is needed that supports the tooling for identifying complex concurrency bugs.  
For example, a visualization is needed to distinguish between the stepping of messages that are exchanged by actors and stepping through the sequential code of each actor.
Ideally, a visualization could also highlight, based on the source code, that certain messages are independent of each other, because there is no direct ordering relationship between them.

\section{Acknowledgments}
This research is funded by a collaboration grant of the Austrian Science Fund (FWF) with the project I2491-N31 and the Research Foundation Flanders (FWO Belgium). 

\begin{landscape}
\section*{Appendix: Table 3 Catalog of Bugs Found in Actor-based Programs}
{\small
\begin{longtable}{p{2.2cm}|p{1.1cm}|p{5.6cm}|p{4.8cm}|p{3.7cm}|p{1.4cm}}%
   \hline
   Bug Type & Id & Bug Pattern & Observable Behavior & Source Reporting the Bug & Language\\ \hline
	
	{Message order violation} & bug-1 & incorrect execution order of two processes when registering a name for a pid in the Process Registry & runtime exception & Fig. 1 in \cite{Christakis:2010} & Erlang \\ \hline
	
	{Memory inconsistency} & bug-2 & insert and write in tables of Erlang Term Storage with public access & inconsistency of values in the tables & Fig. 2 in \cite{Christakis:2010} & Erlang \\ \hline
	
	{Memory inconsistency} & bug-3 & insert and write in tables (dirty operations in Mnesia database) & inconsistency of values in the tables & Fig. 2 in \cite{Christakis:2010} & Erlang \\ \hline
	
	{Communi-cation deadlock} & bug-4 & receive statement with no messages & process in waiting state due to an orphan message & Fig. 1 in \cite{Christakis:2011} & Erlang \\ \hline
		 		
 	{Memory inconsistency} & bug-5 & testing insert operations in parallel (Mnesia database) & exception or inconsistent return values & Sec. 5 of \cite{Hughes:2011} & Erlang \\ \hline
 	
 	{Memory inconsistency} & bug-6 &  testing open\_file in parallel with other operations of dets API (Mnesia database) & inconsistency when visualizing the table's contents & Sec. 5 of \cite{Hughes:2011} & Erlang \\ \hline
 	
 	{Memory inconsistency} & bug-7 & open, close and reopen the file, besides running three processes in parallel (Mnesia database) & integrity checking failed due to premature\_eof error & Sec. 5 of \cite{Hughes:2011} & Erlang \\ \hline
 	
 	{Memory inconsistency} & bug-8 & changes in the dets server state & integrity checking failed (Mnesia database) & Sec. 5 of \cite{Hughes:2011} & Erlang \\ \hline
 	
 	{Communi-cation deadlock} & bug-9 & receive statement with no messages & process in waiting state due to an orphan message (server waits for ping requests) & Program 2 and Test code 2 in \cite{Gotovos:2011} & Erlang \\ \hline
 
	{Communi-cation deadlock} & bug-10 & message sent to a finished process, the finished process exit without replying & process blocks due to an orphan message & Test code 5 in \cite{Gotovos:2011} & Erlang \\ \hline
 	 	
 	{Message order violation} & bug-11 & spawned process that terminates before its Pid is register by the parent process  & process will crash and exits abnormally due to an orphan message & Fig. 1 in \cite{Christakis:2013} & Erlang \\ \hline

	{Bad message interleaving} & bug-12 & actor execute a third message between two consecutive messages &  inconsistent values of variables & Fig. 2 in \cite{Lauterburg:2009} & Actor-Foundry \\ \hline
	
	{Message order violation} & bug-13 & incorrect order of execution of two message receives & the program throws an exception because of a null value &  Listing 1 in \cite{Tasharofi:2013} & Scala \\ \hline
   
   	{Message order violation} & bug-14 & the second message is executed with the value of the first message & actions are performed over the wrong variable & Fig. 4 in \cite{Zheng:2011} & JavaScript\\ \hline
   
   	{Bad message interleaving} & bug-15 & use of a variable not initialized by other methods before it was defined & out of bounds exception & Fig. 4 in \cite{Zheng:2011} & JavaScript\\ \hline
   
   	{Message order violation} & bug-16 & race between HTML parsing and user actions & application crash & Fig. 1 in \cite{Raychev:2013} & JavaScript\\ \hline

	{Message order violation} & bug-17 & race between execution of a script and rendering of an input text box & inconsistency in the value of the variable (storing text the user entered) & Fig. 2 in \cite{Petrov:2012} & JavaScript \\ \hline
		 
    {Message order violation } & bug-18 & race between creation of HTML element and using the element & throw an exception that can lead the application to crash & Fig. 3 in \cite{Petrov:2012} & JavaScript \\ \hline
   
    {Message order violation} & bug-19 & invocation of a function before parsing of the same function & application crash  & Fig. 4 in \cite{Petrov:2012} & JavaScript \\ \hline
   
    {Message order violation} & bug-20 &  iframe's load event fires before the script executes & event handler will never run & Fig. 5 in \cite{Petrov:2012} & JavaScript \\ \hline
 	
	{Bad message interleaving} & bug-21 & execution of an operation (changing the workspace) between two other operations (starting the file transmission and the completion of the transmission) & exception of variable undefined & Fig. 6 in \cite{Hong:2014} & JavaScript\\ \hline
	
 	{Bad message interleaving} & bug-22 & event handler updates DOM between two input events that manipulate the same DOM element & error because of a null value & Fig.3 in \cite{Hong:2014} & JavaScript\\ \hline
   		
 	{Message order violation} & bug-23 & user input invokes a function before it has been defined/loaded & application crashes (due to unexpected turn termination) & Fig. 2 in \cite{Hong:2014} & JavaScript\\ \hline

\caption{Catalog of bugs found in actor-based programs}
\label{tab:bugs}
\end{longtable}
}%
\end{landscape}

\bibliographystyle{splncs03}
%\bibliography{references}

\end{document}